\documentclass[twocolumn]{aastex7}

\def\lapp{\ifmmode\stackrel{<}{_{\sim}}\else$\stackrel{<}{_{\sim}}$\fi}
\def\gapp{\ifmmode\stackrel{>}{_{\sim}}\else$\stackrel{>}{_{\sim}}$\fi}
\usepackage{graphicx}
\usepackage{multirow}
\usepackage{color}
\usepackage{amsmath}
\usepackage{soul}
\usepackage{hyperref}
\newcommand{\fluxcgs}{\ensuremath{\mathrm{erg}\,\mathrm{s}^{-1}\,\mathrm{cm}^{-2}}}

\def\footnoterule{\hrule width 0.3\textwidth \vspace*{0.1cm}}
\shorttitle{}
\shortauthors{}
\begin{document}

\title{Multi-wavelength Study of HESS~J0632+057: New Insights into Pulsar-Disk Interaction}

\correspondingauthor{Hongjun An}
\email{hjan@cbnu.ac.kr}
\author[0000-0002-9103-506X]{Jaegeun Park}
\email{}
\affiliation{Department of Astronomy and Space Science, Chungbuk National University, Cheongju, 28644, Republic of Korea}
\author[0000-0002-6389-9012]{Hongjun An}
\email{hjan@cbnu.ac.kr}
\affiliation{Department of Astronomy and Space Science, Chungbuk National University, Cheongju, 28644, Republic of Korea}
\author[0000-0003-0226-9524]{Chanho Kim}
\email{}
\affiliation{Department of Astronomy and Space Science, Chungbuk National University, Cheongju, 28644, Republic of Korea}
\author[0000-0002-2741-2959]{Natalie Matchett}
\email{}
\affiliation{Department of Physics, University of the Free State, 205 Nelson Mandela Dr., Bloemfontein, 9300, South Africa}
\author[0000-0002-9709-5389]{Kaya Mori}
\email{}
\affiliation{Columbia Astrophysics Laboratory, 550 West 120th Street, New York, NY 10027, USA}
\author[0000-0003-1873-7855]{Brian van Soelen}
\email{}
\affiliation{Department of Physics, University of the Free State, 205 Nelson Mandela Dr., Bloemfontein, 9300, South Africa}

\nocollaboration{all}

\author{A.~Archer}\email{}\affiliation{Department of Physics and Astronomy, DePauw University, Greencastle, IN 46135-0037, USA}
\author[0000-0002-3886-3739]{P.~Bangale}\email{}\affiliation{Department of Physics and Astronomy and the Bartol Research Institute, University of Delaware, Newark, DE 19716, USA}
\author[0000-0002-9675-7328]{J.~T.~Bartkoske}\email{}\affiliation{Department of Physics and Astronomy, University of Utah, Salt Lake City, UT 84112, USA}
\author[0000-0003-2098-170X]{W.~Benbow}\email{}\affiliation{Center for Astrophysics $|$ Harvard \& Smithsonian, Cambridge, MA 02138, USA}
\author[0000-0001-6391-9661]{J.~H.~Buckley}\email{}\affiliation{Department of Physics, Washington University, St. Louis, MO 63130, USA}
\author[0009-0001-5719-936X]{Y.~Chen}\email{}\affiliation{Department of Physics and Astronomy, University of California, Los Angeles, CA 90095, USA}
\author{A.~J.~Chromey}\email{}\affiliation{Center for Astrophysics $|$ Harvard \& Smithsonian, Cambridge, MA 02138, USA}
\author[0000-0003-1716-4119]{A.~Duerr}\email{}\affiliation{Department of Physics and Astronomy, University of Utah, Salt Lake City, UT 84112, USA}
\author[0000-0002-1853-863X]{M.~Errando}\email{}\affiliation{Department of Physics, Washington University, St. Louis, MO 63130, USA}
\author{M.~Escobar~Godoy}\email{}\affiliation{Santa Cruz Institute for Particle Physics and Department of Physics, University of California, Santa Cruz, CA 95064, USA}
\author[0000-0002-5068-7344]{A.~Falcone}\email{}\affiliation{Department of Astronomy and Astrophysics, 525 Davey Lab, Pennsylvania State University, University Park, PA 16802, USA}
\author{S.~Feldman}\email{}\affiliation{Department of Physics and Astronomy, University of California, Los Angeles, CA 90095, USA}
\author[0000-0001-6674-4238]{Q.~Feng}\email{}\affiliation{Department of Physics and Astronomy, University of Utah, Salt Lake City, UT 84112, USA}
\author[0000-0002-2636-4756]{S.~Filbert}\email{}\affiliation{Department of Physics and Astronomy, University of Utah, Salt Lake City, UT 84112, USA}
\author[0000-0002-1067-8558]{L.~Fortson}\email{}\affiliation{School of Physics and Astronomy, University of Minnesota, Minneapolis, MN 55455, USA}
\author[0000-0003-1614-1273]{A.~Furniss}\email{}\affiliation{Santa Cruz Institute for Particle Physics and Department of Physics, University of California, Santa Cruz, CA 95064, USA}
\author[0000-0002-0109-4737]{W.~Hanlon}\email{}\affiliation{Center for Astrophysics $|$ Harvard \& Smithsonian, Cambridge, MA 02138, USA}
\author[0000-0003-3878-1677]{O.~Hervet}\email{}\affiliation{Santa Cruz Institute for Particle Physics and Department of Physics, University of California, Santa Cruz, CA 95064, USA}
\author[0000-0001-6951-2299]{C.~E.~Hinrichs}\email{}\affiliation{Center for Astrophysics $|$ Harvard \& Smithsonian, Cambridge, MA 02138, USA and Department of Physics and Astronomy, Dartmouth College, 6127 Wilder Laboratory, Hanover, NH 03755 USA}
\author[0000-0002-6833-0474]{J.~Holder}\email{}\affiliation{Department of Physics and Astronomy and the Bartol Research Institute, University of Delaware, Newark, DE 19716, USA}
\author[0000-0002-1432-7771]{T.~B.~Humensky}\email{}\affiliation{Department of Physics, University of Maryland, College Park, MD, USA and NASA GSFC, Greenbelt, MD 20771, USA}
\author[0000-0002-1089-1754]{W.~Jin}\email{}\affiliation{Department of Physics and Astronomy, University of California, Los Angeles, CA 90095, USA}
\author[0009-0008-2688-0815]{M.~N.~Johnson}\email{}\affiliation{Santa Cruz Institute for Particle Physics and Department of Physics, University of California, Santa Cruz, CA 95064, USA}
\author[0000-0002-3638-0637]{P.~Kaaret}\email{}\affiliation{Department of Physics and Astronomy, University of Iowa, Van Allen Hall, Iowa City, IA 52242, USA}
\author{M.~Kertzman}\email{}\affiliation{Department of Physics and Astronomy, DePauw University, Greencastle, IN 46135-0037, USA}
\author{M.~Kherlakian}\email{}\affiliation{Fakult\"at f\"ur Physik \& Astronomie, Ruhr-Universit\"at Bochum, D-44780 Bochum, Germany}
\author[0000-0003-4785-0101]{D.~Kieda}\email{}\affiliation{Department of Physics and Astronomy, University of Utah, Salt Lake City, UT 84112, USA}
\author[0000-0002-4260-9186]{T.~K.~Kleiner}\email{}\affiliation{DESY, Platanenallee 6, 15738 Zeuthen, Germany}
\author[0000-0002-4289-7106]{N.~Korzoun}\email{}\affiliation{Department of Physics and Astronomy and the Bartol Research Institute, University of Delaware, Newark, DE 19716, USA}
\author[0000-0002-5167-1221]{S.~Kumar}\email{}\affiliation{Department of Physics, University of Maryland, College Park, MD, USA }
\author[0000-0003-4641-4201]{M.~J.~Lang}\email{}\affiliation{School of Natural Sciences, University of Galway, University Road, Galway, H91 TK33, Ireland}
\author[0000-0003-3802-1619]{M.~Lundy}\email{}\affiliation{Physics Department, McGill University, Montreal, QC H3A 2T8, Canada}
\author[0000-0001-9868-4700]{G.~Maier}\email{gernot.maier@desy.de}\affiliation{DESY, Platanenallee 6, 15738 Zeuthen, Germany}
\author[0000-0002-1499-2667]{P.~Moriarty}\email{}\affiliation{School of Natural Sciences, University of Galway, University Road, Galway, H91 TK33, Ireland}
\author[0000-0002-3223-0754]{R.~Mukherjee}\email{}\affiliation{Department of Physics and Astronomy, Barnard College, Columbia University, NY 10027, USA}
\author{M.~Ohishi}\email{}\affiliation{Institute for Cosmic Ray Research, University of Tokyo, 5-1-5, Kashiwa-no-ha, Kashiwa, Chiba 277-8582, Japan}
\author[0000-0002-4837-5253]{R.~A.~Ong}\email{}\affiliation{Department of Physics and Astronomy, University of California, Los Angeles, CA 90095, USA}
\author[0000-0003-3820-0887]{A.~Pandey}\email{}\affiliation{Department of Physics and Astronomy, University of Utah, Salt Lake City, UT 84112, USA}
\author[0000-0001-7861-1707]{M.~Pohl}\email{}\affiliation{Institute of Physics and Astronomy, University of Potsdam, 14476 Potsdam-Golm, Germany and DESY, Platanenallee 6, 15738 Zeuthen, Germany}
\author[0000-0002-0529-1973]{E.~Pueschel}\email{}\affiliation{Fakult\"at f\"ur Physik \& Astronomie, Ruhr-Universit\"at Bochum, D-44780 Bochum, Germany}
\author[0000-0002-4855-2694]{J.~Quinn}\email{}\affiliation{School of Physics, University College Dublin, Belfield, Dublin 4, Ireland}
\author{P.~L.~Rabinowitz}\email{}\affiliation{Department of Physics, Washington University, St. Louis, MO 63130, USA}
\author[0000-0002-5351-3323]{K.~Ragan}\email{}\affiliation{Physics Department, McGill University, Montreal, QC H3A 2T8, Canada}
\author[0000-0002-7523-7366]{D.~Ribeiro}\email{}\affiliation{School of Physics and Astronomy, University of Minnesota, Minneapolis, MN 55455, USA}
\author{E.~Roache}\email{}\affiliation{Center for Astrophysics $|$ Harvard \& Smithsonian, Cambridge, MA 02138, USA}
\author[0000-0003-1387-8915]{I.~Sadeh}\email{}\affiliation{DESY, Platanenallee 6, 15738 Zeuthen, Germany}
\author[0000-0002-3171-5039]{L.~Saha}\email{}\affiliation{Center for Astrophysics $|$ Harvard \& Smithsonian, Cambridge, MA 02138, USA}
\author{G.~H.~Sembroski}\email{}\affiliation{Department of Physics and Astronomy, Purdue University, West Lafayette, IN 47907, USA}
\author[0000-0002-9856-989X]{R.~Shang}\email{}\affiliation{Department of Physics and Astronomy, Barnard College, Columbia University, NY 10027, USA}
\author{J.~V.~Tucci}\email{}\affiliation{Department of Physics, Indiana University Indianapolis, Indianapolis, Indiana 46202, USA}
\author{V.~V.~Vassiliev}\email{}\affiliation{Department of Physics and Astronomy, University of California, Los Angeles, CA 90095, USA}
\author{A.~Weinstein}\email{}\affiliation{Department of Physics and Astronomy, Iowa State University, Ames, IA 50011, USA}
\author[0000-0003-2740-9714]{D.~A.~Williams}\email{}\affiliation{Santa Cruz Institute for Particle Physics and Department of Physics, University of California, Santa Cruz, CA 95064, USA}
\author[0000-0002-2730-2733]{S.~L.~Wong}\email{}\affiliation{Physics Department, McGill University, Montreal, QC H3A 2T8, Canada}
\author{T.~Yoshikoshi}\email{}\affiliation{Institute for Cosmic Ray Research, University of Tokyo, 5-1-5, Kashiwa-no-ha, Kashiwa, Chiba 277-8582, Japan}

\collaboration{all}{(VERITAS Collaboration)}

\begin{abstract}
We present an analysis of new multi-wavelength observations of the TeV gamma-ray binary HESS~J0632+057,
conducted using SALT, Swift, NuSTAR, and VERITAS in 2023--2024.
By combining these new data with archival observations, we confirm previous suggestions of orbital
variability in the source's X-ray spectrum, including increased X-ray absorption at the orbital
phase interval of $\phi\approx0.3\textrm{--}0.4$. The source's X-ray flux within this phase interval seems to have exhibited a significant change on an orbital timescale.
Additionally, occasional short-term variations in the X-ray band on a timescale of less than 3\,days have been observed. 
The measured duration of the increased absorbing column density and the flux variability timescales can provide
clues about the interaction between the putative pulsar and the Be companion's disk
if, as previously suggested, the pulsar crosses
the disk at this phase interval. Moreover, the new contemporaneous X-ray and TeV observations
around the pulsar-crossing phases revealed independent variability in the X-ray and TeV fluxes,
contrary to a previous observation of concurrent flux increases.
While these observations alone cannot provide definitive conclusions,
we discuss our results in the context of pulsar-disk interaction and intrabinary shock emission scenarios.
\end{abstract}

\bigskip 
\section{Introduction}
\label{sec:intro}
TeV gamma-ray binaries (TGBs), composed of a compact object (neutron star or black hole)
and a massive stellar companion, exhibit diverse high-energy phenomena
driven by interactions between the two components. These sources are intriguing
due to their ability to accelerate particles to very high energies.
There are primarily two models proposed for TGBs: the microquasar and pulsar models \citep[e.g.,][]{m12}.
The microquasar model assumes that relativistic jets of a putative black hole accelerate
particles \citep[e.g.,][]{brp06}, while the pulsar model proposes an intrabinary shock (IBS)
formed by the collision of the pulsar and companion winds as the driver for
accelerating pulsar-wind particles \citep[e.g.,][]{Dubus2013}.
The key factor distinguishing these scenarios is the nature of the compact object.
Although pulsars have been definitively identified in only three TGBs \citep[PSR~B1259$-$63, PSR~J2032+4127, and LS~I~+61$^\circ$~303;][respectively]{Johnston+1992,J2032VER2018,Weng+2022} out of nine known systems,
evidence suggests that neutron stars may power other TGBs \citep[e.g.,][]{Dubus2013,An2015}, including HESS~J0632+057 \citep[e.g.,][]{Moritani+2018}.

The presence of a decretion disk around a Be-type companion star in some TGBs further complicates and
enriches the observed high-energy phenomena. PSR~B1259$-$63 serves as an archetypal example \citep[][]{Johnston+1992}.
Here, the pulsar is thought to interact with the Be star's disk during specific orbital phases,
triggering complex multi-band phenomena like flares, spectral variations \citep[e.g.,][]{Chernyakova2024},
and ejection of matter from the system \citep[e.g.,][]{Pavlov+2015}.
Although these phenomena are likely connected to this interaction, our understanding
of the underlying physics remains incomplete.

\newcommand{\marka}{\tablenotemark{\tiny{\rm a}}}
\newcommand{\markb}{\tablenotemark{\tiny{\rm b}}}
\newcommand{\markc}{\tablenotemark{\tiny{\rm c}}}
\begin{table*}[t]
\vspace{-0.0in}
\begin{center}
\caption{X-ray data used in this study}
\label{ta:ta1}
\vspace{-0.05in}
\scriptsize{
\begin{tabular}{lcccc} \hline
Observatory &  Obs. ID & Date  & Total exposure & $N_{\rm obs}$ \\ \hline
&   & (MJD) & (ks) &  \\ \hline
Swift & 00015903001--00097195011 & 54857--60372 & 1018 & 289 \\
NuSTAR & 30362001002--91002305004 & 58079--60353 & 269 & 8 \\
Chandra & 13237--20975\marka & 55606--58170 & 192 & 5 \\
XMM-Newton & 0505200101, 0821370201 & 54360, 58372 & 49 & 2 \\
Suzaku & 403018010, 404027010 & 54580, 54942  &104 &  2 \\ \hline
\end{tabular}}
\end{center}
\vspace{-0.5 mm}
\footnotesize{
\marka{Data contained in\dataset[Chandra Data Collection (CDC) 345]{https://doi.org/10.25574/cdc.345}.}}
\end{table*}
 
Theoretical and observational studies have extensively investigated physical processes
related to the enhanced high-energy emission observed during or around the
disk-interaction phase of PSR~B1259$-$63 \citep[e.g.,][]{Chernyakova2024,Khangulyan+07}.
The detection of X-ray polarization from PSR~B1259$-$63 by IXPE \citep[][]{Kaaret2024} confirms that its X-ray emission originates from synchrotron radiation produced by electrons interacting with the IBS magnetic field, which is oriented perpendicular to the shock cone axis.
The Be companion's substantial equatorial decretion disk may compress
the pulsar-wind shock, increasing the magnetic field $B$ within the IBS
and consequently the X-ray emission from it as proposed by \citet{TavaniArons1997}.
Dense seed-photon fields of the companion and disk at the interaction phase
can give rise to high TeV flux through inverse-Compton (IC) upscattering.

Investigations of high-energy phenomena related to the pulsar-disk interaction within
TGBs have thus far focused on three systems hosting Be companions: PSR~B1259$-$63, LS~I~$+61^\circ$303, and PSR~J2023+4127. Incorporating additional systems into
such studies is crucial for advancing
our understanding of the underlying interaction physics.
The TGB HESS~J0632+057 \citep[][]{HESSJ06372007}, which hosts a Be-type companion \citep[HD 259440;][]{Aragona+2010} orbiting with a period of 310--320 days, emerges as a promising candidate for further study of pulsar-disk interactions.
Although the nature of the compact object remains uncertain, constraints on its mass \citep[][]{Moritani+2018}
and X-ray image analysis \citep[][]{Kargaltsev+2022} suggest a neutron star.
Notably, HESS~J0632+057 (J0632 hereafter) exhibits X-ray and TeV flares
around the orbital phase $\phi=0.35$ \citep[][]{Aliu2014},
interpreted as a signature of pulsar-disk interaction.
Furthermore, previous studies \citep[e.g.,][]{Moritani+2018,Malyshev+2019} suggest
a higher hydrogen column density ($N_{\rm H}$)
at this phase, further supporting the disk interaction hypothesis.
These previous studies utilized different orbital periods
(e.g., $P_B=310$--$320$\,days), although their conclusions remain relatively
insensitive to the precise value of $P_B$.

\citet{Tokayer+2021} carried out a multi-wavelength study, combining contemporaneous
optical, X-ray, and TeV observations of J0632. They focused on the phase interval
of the second X-ray peak ($\phi\sim 0.5$--$0.8$) and demonstrated
significant variability in the X-ray photon index. In contrast, no
significant variation was observed in the equivalent widths (EWs) of the H\,$\alpha$ and H\,$\beta$ emission lines,{\footnote{For simplicity, we report the EW values for emission lines as the absolute value.} implying insignificant pulsar-disk interaction.
Furthermore, they modeled the X-ray and TeV spectral energy distributions (SEDs)
using a one-zone IBS scenario to infer the system properties during this phase interval.
Intriguingly, they also found evidence for a $N_{\rm H}$ increase
around $\phi\sim0.13$, suggesting
that pulsar-disk interaction might also occur in that phase interval.

While \citet{Tokayer+2021} characterized the
multi-wavelength properties of J0632 during the orbital phase interval $\phi\sim 0.5$--$0.8$,
a comprehensive understanding of the source requires contemporaneous
multi-wavelength data in other phases.
The X-ray/TeV flux enhancements and spectral changes observed during $\phi\sim0.3\textrm{--}0.4$ hold significant promise for unraveling the physics of the pulsar-disk interaction. Variability in the emissions from the companion and its disk can also provide crucial insights into the origin of TeV emission since the IR and optical emissions serve as seed photons for IC scattering.
To better characterize these features and explore potential explanations,
we conducted a targeted multi-wavelength observation campaign of J0632
around $\phi\sim0.3\textrm{--}0.4$.

This paper presents our analysis of X-ray (Section~\ref{sec:sec2}), optical (Section~\ref{sec:opt}), and TeV data (Section~\ref{sec:sec3}) obtained between 2023 December and 2024 March. To achieve a comprehensive X-ray characterization across the orbit, we also analyze archival X-ray observations. Following the data analysis, we discuss our findings in the context
of IBS emission and pulsar-disk interaction (Section~\ref{sec:sec4}),
and conclude with a summary in Section~\ref{sec:sec5}.

\section{X-ray Data Analysis}
\label{sec:sec2}
Our X-ray observation campaign for J0632, performed by Swift/XRT and NuSTAR,
collected data in the orbital phase range around $\phi\approx0.35$, between 2023 December and 2024 March.
To complement these new data, we also re-analyzed archival X-ray observations, encompassing all orbital phases, to obtain
high-quality light curves and phase-resolved spectra. Details of both the new and archival
data used in this work are provided in Table~\ref{ta:ta1}.

While many of these observations have already been reported
\citep[e.g.,][]{Malyshev+2019,Tokayer+2021,Kargaltsev+2022},
this work incorporates 26 new observations (23 Swift and 3 NuSTAR) to provide the most
comprehensive analysis of the source's X-ray emission to date.
We use the 0.5--10\,keV band for Swift/XRT, Chandra, XMM-Newton, and Suzaku data analysis, and the 3--20\,keV band for NuSTAR analysis. We verified that slightly changing the energy band (e.g., 0.3--10\,keV for XRT) does not significantly alter the results.

\subsection{Data reduction}
\label{sec:sec2_1}

\begin{figure*}
\centering
\begin{tabular}{cc}
\hspace{-7mm}
\includegraphics[width=3.5 in]{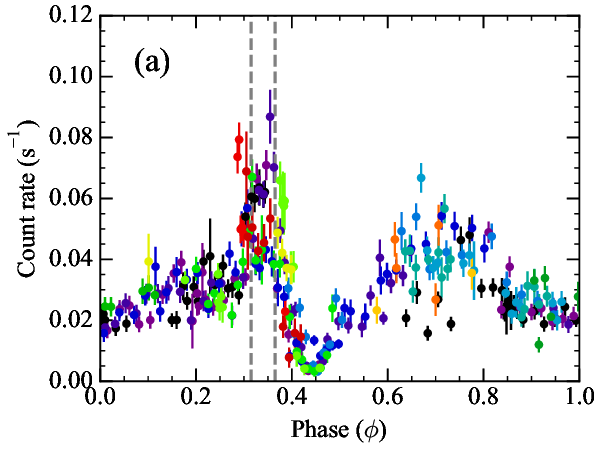} &
\hspace{-7mm}
\includegraphics[width=3.5 in]{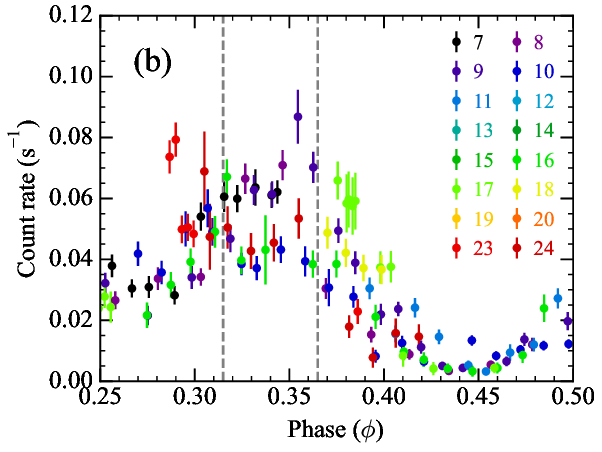} \\
\hspace{-7mm}
\includegraphics[width=3.43 in]{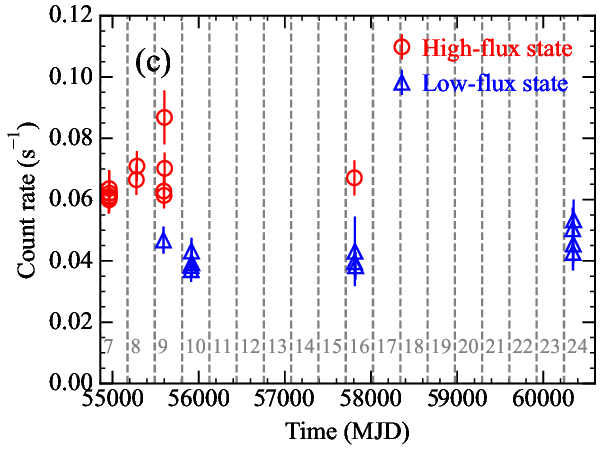} &
\includegraphics[width=3.5 in]{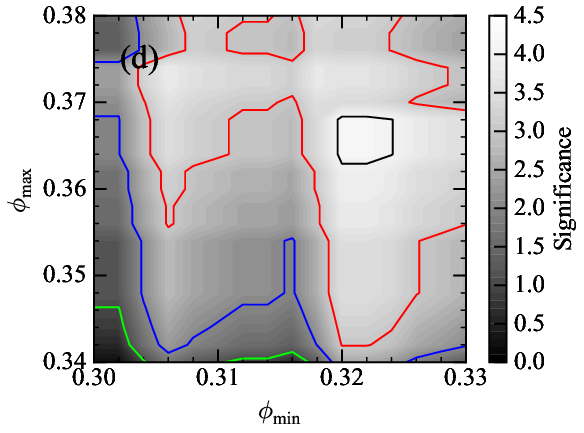} \\
\end{tabular}
\figcaption{The 0.5--10\,keV light curves of J0632 measured by Swift/XRT (a--c).
(a) Light curve folded on the orbital period $P_B=316.65$\,days.
(b) Zoomed-in view of the folded light curve near the X-ray peak,
highlighting the phase interval $\phi=0.315$--$0.365$ (marked by vertical dashed lines in panels a and b) corresponding to the previously-suggested pulsar-disk interaction phase.
Different colors represent individual orbital cycles.
(c) Unfolded light curve for the phase interval $\phi=0.315$--$0.365$, with vertical
lines indicating orbital cycles (gray numbers). 
(d) Probabilities that the light-curve structure in panel (c) arises from random variation
as a function of phase-interval selection ($\phi_{\rm min}$--$\phi_{\rm max}$). The green, blue, red, and black contours correspond to significance levels of 1, 2, 3, and 4, respectively.
\label{fig:fig1}}
\vspace{0mm}
\end{figure*}

We followed the standard data analysis procedures recommended by each observatory
to reduce the observational data.
The Swift/XRT data, obtained with the photon counting (PC) mode, were reprocessed with the {\tt xrtpipeline} script
along with the calibration database version 20240522.
Source and background events were extracted within a circular region of radius
$R=30''$ and an annular region of radii $R=50''\textrm{--}100''$, respectively.
The Chandra data were reprocessed with the {\tt chandra\_repro} script, and we used
a $R=3''$ circle and a $R=5''$--$10''$ annular region for the source and background extraction, respectively.
One Chandra observation (Obsid. 13237) was made utilizing the continuous-clock mode. For this
observation, we extracted source events within a rectangular region with a width of $5''$
and background events from two rectangular regions with widths of $10''$.
The XMM-Newton data were processed using the {\tt emproc} and {\tt epproc} scripts within XMM-SAS v20230412\_1735. Source and background spectra were extracted from circular regions with radii of $20''$ and $40''$, respectively.
The Suzaku data were reprocessed with the {\tt aepipeline} tool, and spectra were
extracted within a $R=1.5'$ circle and a $R=1.5'$--$3'$ annulus for the source and background, respectively.
Initial reduction of the NuSTAR data was performed with the {\tt nupipeline} tool,
and we used $R=60''$ for the source extraction. Because of the non-uniform background in the NuSTAR data caused
by stray light, we used the {\tt nuskybgd} simulations \citep[][]{Wik2014} to estimate the background.

Given the low source brightness and observed count rate ($<0.1$\,cps; see Sections~\ref{sec:sec2_1}--\ref{sec:sec2_4}), photon pile-up \citep[Swift/XRT, Suzaku, XMM-Newton, Chandra; see also][]{Kargaltsev+2022} and deadtime (NuSTAR) effects are negligible.

\subsection{Timing analysis}
\label{sec:sec2_2}
To establish a robust orbital ephemeris for J0632,
we constructed an exposure-corrected Swift/XRT light curve binned in counts
after barycenter-correcting event arrival times using a position of  (RA, Decl)=($98.2469^\circ$, $5.8003^\circ$) for J0632. Data from the other X-ray observatories are not included in our initial analysis due to their distinct spectral responses.

Folding the new XRT data on the previously reported period of $P_B=317.3$\,days \citep[][]{Tokayer+2021}
revealed a misalignment with existing data.
To refine the period, we performed a timing analysis with the count-rate measurements
utilizing the $H$ test \citep[][]{drs89}.
This analysis resulted in a best-fit period of $316.65\pm0.31$\,days,
statistically consistent with prior reports of \citet{Malyshev+2019} and \citet{Adams+2021}.
To investigate the impact of data from the other observatories, we incorporated their data by converting the inferred spectra (Section~\ref{sec:sec2_3}) to equivalent Swift count rates. This analysis resulted in a period of $P_B=316.91\pm0.28$\,days, which is statistically consistent with the result obtained using Swift data alone.
We adopt $P_B=316.65$\,days and the reference epoch
$T_0=$54857 MJD \citep[][]{Bongiorno+11} for subsequent analysis. Orbital cycle numbers are defined as in \citet{Adams+2021}, with cycle 1 starting at MJD~52953. We verified that alternative
choices of $P_B$ within the 1$\sigma$ uncertainty do not significantly
impact the results presented below.

Figure~\ref{fig:fig1}a presents the source's 0.5--10\,keV light curve,
folded on our $P_B$. The multi-orbit data (color-coded for clarity according to the orbital cycles) align well, confirming the previously reported features: a peak near $\phi\approx 0.35$, a dip around
$\phi\approx0.45$, and a broad bump between phases 0.6 and 0.8 \citep[][]{Aliu2014}.
The light curve shows strong variability, and we found a shortest timescale of 1\,day
for the variability at the 3$\sigma$ confidence level. This timescale is limited by the cadence of Swift observations. We note that \citet{Kargaltsev+2022} have reported flux variations in this source on timescales shorter than one day.

Notably, the light curve reveals complex variability within the phase interval around the X-ray peak at $\phi\approx 0.35$ (Figure~\ref{fig:fig1}b).
Within the pulsar-disk interaction phase, $\phi\approx0.31\textrm{--}0.37$, the flux displays pronounced fluctuations, characterized by intermittent high-flux events (flares) superimposed on a low flux level.
To further examine this variability, we generated an unfolded light curve using 24 measurements within $\phi=0.315\textrm{--}0.365$, presented in Figure~\ref{fig:fig1}c. 
This light curve reveals a distinct temporal pattern, with higher (than average) count rates observed prior to MJD 55800 compared to subsequent measurements, indicating variability on an orbital timescale. Additionally, the source displays short-term transitions between flux levels within individual orbital cycles, with a timescale of $<3$\,days.

To assess the significance of this pattern,
we performed a Wald-Wolfowitz runs test \citep[][]{Wald1940}, evaluating the probability that this pattern of low and high states resulted from random fluctuations in count rates drawn from a single distribution. The resulting probability, $p\lapp 0.003$, strongly indicates that the observed pattern is attributable to long-term flux modulation rather than stochastic variations. This conclusion was verified to be robust across a range of phase selection intervals ($\phi_{\rm min}$ to $\phi_{\rm max}$; Figure~\ref{fig:fig1}d).

Motivated by this temporal pattern, we explored whether the observations within the interaction phase exhibit flux grouping independent of the temporal ordering. Utilizing the gap statistic \citep[][]{Tibshirani2001}, we identified an optimal grouping of two clusters.
However, the corresponding gap-statistic value is relatively low, and simulations of uniformly distributed data yielded a chance probability of $0.2$ of a similar two-cluster formation, indicating that the observed clustering is not statistically robust. Therefore, while we utilize the terms ``low-flux'' and ``high-flux states'' for descriptive convenience, it is crucial to recognize that the flux distribution within this phase may be continuous (see also Section~\ref{sec:sec4_2}).

\begin{figure*}
\centering
\begin{tabular}{ccc}
\hspace{-5mm}
\includegraphics[width=2.35 in]{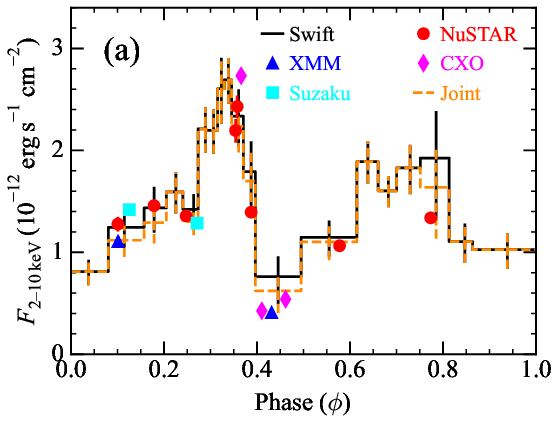} &
\hspace{-5mm}
\includegraphics[width=2.44 in]{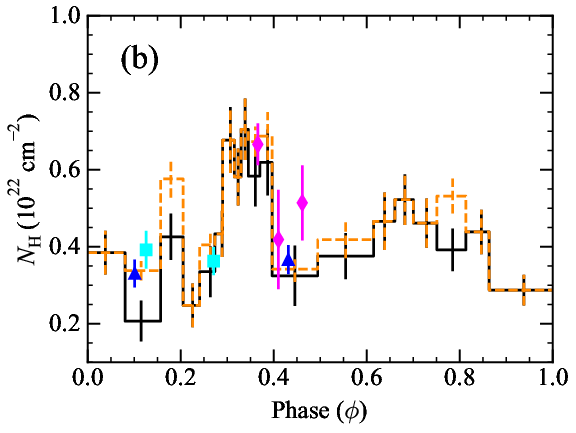} &
\hspace{-5mm}
\includegraphics[width=2.42 in]{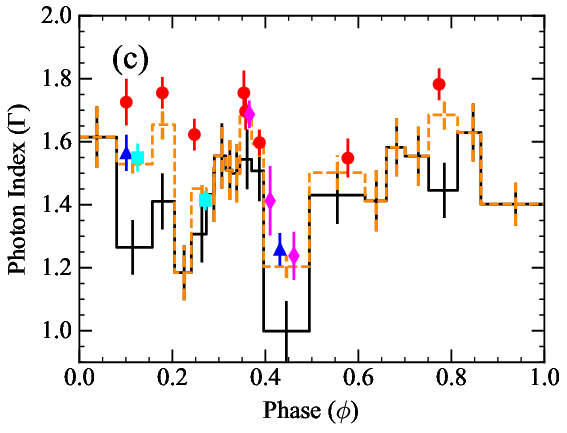} \\
\end{tabular}
\figcaption{Plots showing (a) the 2--10\,keV flux ($F_{2-10\rm keV}$), (b) $N_{\rm H}$, and (c) $\Gamma$ inferred from PL fits (Section~\ref{sec:sec2_3}).
Measurements from Swift/XRT, Chandra, XMM, Suzaku, and NuSTAR are shown as black lines, pink diamonds,
blue triangles, cyan squares, and red circles, respectively. The orange dashed line indicates
parameter values obtained from joint fits of the multi-instrument data.
\label{fig:fig2}}
\vspace{0mm}
\end{figure*}

\subsection{Phase-resolved Spectral Analysis}
\label{sec:sec2_3}
To investigate the orbital variability in the source's emission,
we performed a phase-resolved spectral analysis.
To achieve a meaningful characterization of the spectral variations
with the high-cadence Swift/XRT data, we ensured that each phase bin contains
$>$1000 source counts in the 0.5--10\,keV XRT data
and that the binning does not significantly blur important features in the light curve (Figure~\ref{fig:fig1}a).
We extracted source and background spectra using the regions described
in Section~\ref{sec:sec2_1} and generated appropriate response files for each observatory using the standard tools:
{\tt xrtmkarf} for XRT,\footnote{https://www.swift.ac.uk/analysis/xrt/rmfarf.php}
{\tt nuproduct} for NuSTAR, {\tt rmfgen} and {\tt arfgen} for XMM-Newton,
{\tt specextract} for Chandra, and {\tt XSELECT}
for Suzaku.\footnote{http://www.hep.nsdc.cn/browsall/suzaku/doc/general/suzak\\u\_abc\_guide\_5.0.pdf}

We grouped the XRT spectra such that each spectral bin contained at least one count, and employed the C-statistic \citep{Cash79} for spectral fitting. The spectra from the other instruments were grouped to have a minimum of 30 counts per bin and fit using the
$\chi^2$ statistic. We fit these spectra with an absorbed power-law (PL) model within {\it XSPEC} v12.14.0, using the {\tt tbabs} model with the {\tt wilm} abundances \citep[][]{Wilms2000} and {\tt vern} cross-sections \citep[][]{vfky96} to account for Galactic absorption.
We note that spectral parameters for XRT spectra within the same phase bin
are tied to common values, while parameters for spectra of the other observatories are
independently fit at this stage.
Due to limitations in the low-energy (below 3\,keV) sensitivity of NuSTAR,
$N_{\rm H}$ was fixed at the value derived from the XRT data.
Our fitting results are broadly consistent with previous reports \citep[e.g.,][]{Malyshev+2019}, with the exception of the $N_{\rm H}$ values. We suspect that this discrepancy arose from the distinct abundance tables used in the analyses; by employing the {\tt anrg} abundances \citep[][]{angr89}, we were able to reproduce these previous results.
We present the results in Figure~\ref{fig:fig2},
where the black solid line represents the XRT measurements and the color points show the results
from the other observatories.

The X-ray flux, as expected, shows a similar trend to the light curve,
with a peak, dip, and bump structure.
Our analysis confirms the previously reported variability in both
$N_{\rm H}$ and the photon index $\Gamma$ \citep[e.g.,][]{Malyshev+2019},
as shown in Figure~\ref{fig:fig2}b and c.
The $N_{\rm H}$ trend exhibits a potential double-bump structure with maxima around $\phi\approx0.35$ and 0.7,
although the latter bump is much weaker and requires further investigation by deep (and single-epoch) X-ray data.
The $\Gamma$ values measured by XRT also show variability across the orbital phases;
most notably a prominent spectral hardening coincides with the flux dip at $\phi\approx 0.45$.

Measurements from other soft-band instruments (Chandra, XMM, and Suzaku)
often differ from the Swift XRT results derived from analysis of multi-orbit data. 
While orbit-to-orbit variability in the XRT data could contribute to these differences, they are statistically insignificant and accounted for by parameter covariance. The multi-instrument spectra within each phase bin are well described by a single PL (Table~\ref{ta:ta2}).

NuSTAR fits of the data generally yielded softer $\Gamma$ than those derived from the soft-band instruments (Figure~\ref{fig:fig2}c).
Joint fitting of NuSTAR and soft-band data during phases with a significant $\Gamma$ difference
(e.g., $\phi\sim0.25$; Figure~\ref{fig:fig2}c) indicates that single PL models suffice to explain
the broadband X-ray data. Combining the NuSTAR and soft-band data effectively
resulted in higher $N_{\rm H}$ and $\Gamma$ values compared to fits of the soft-band data alone.
The PL models provided statistically acceptable fits even when $N_{\rm H}$ was fixed at the Swift-measured values.
However, F-test comparisons between PL and broken power-law (BPL) models, with the
latter having a break at 3--5\,keV and $\Delta \Gamma\approx 0.3\textrm{--}0.5$, suggest
that the BPL model is statistically favored
with F-test probabilities of $2\times 10^{-4}$ and $2\times 10^{-5}$ for $\phi=0.2$ and 0.25, respectively.
These probabilities were lower when $N_{\rm H}$ was fixed.
While this might suggest a curved spectral shape, a definitive conclusion is precluded by
known orbit-to-orbit variability in the source's emission.
Therefore, simultaneous observations across a broad X-ray band are warranted for confirmation.

\begin{table}[t]
\vspace{-0.0in}
\begin{center}
\caption{Results of joint fit of multi-instrument data}
\label{ta:ta2}
\vspace{-0.05in}
\scriptsize{
\begin{tabular}{lcccc} \hline
Phase &  $N_{\rm H}$ & $\Gamma$ & $F_{2-10\rm \ keV}$\marka & $\chi^2$/dof \\ \hline
  &   $ (10^{22}\rm \ cm^{-2}) $ &  &  &   \\ \hline
0.000--0.076 & 0.38(6) & 1.61(10) & 0.81(12) & 915.8\,(915) \\
0.076--0.154 & 0.34(3) & 1.53(3) & 1.12(15) & 1150.4\,(1267) \\
0.154--0.204 & 0.58(5) & 1.65(5) & 1.29(19) & 999.8\,(1101) \\
0.204--0.247 & 0.25(6) & 1.18(9) & 1.59(20) & 797.7\,(900) \\
0.247--0.281 & 0.40(3) & 1.45(3) & 1.30(13) & 1421.2\,(1482) \\
0.281--0.298 & 0.43(6) & 1.43(9) & 2.21(22) & 789.4\,(919) \\
0.298--0.316 & 0.68(9) & 1.55(10) & 2.20(22) & 755.4\,(826) \\
0.316--0.331 & 0.58(7) & 1.51(9) & 2.61(32) & 814.0\,(906) \\
0.331--0.346 & 0.71(8) & 1.50(9) & 2.69(22) & 827.0\,(946) \\
0.346--0.376 & 0.67(4) & 1.69(3) & 2.24(24) & 1076.2\,(1164) \\
0.376--0.399 & 0.69(6) & 1.59(4) & 1.70(31) & 1013.7\,(1089) \\
0.399--0.492 & 0.34(3) & 1.20(4) & 0.62(19) & 1197.8\,(1223) \\
0.492--0.618 & 0.42(5) & 1.50(5) & 1.10(17) & 954.7\,(1017) \\
0.618--0.659 & 0.47(7) & 1.41(10) & 1.89(21) & 810.4\,(885) \\
0.659--0.705 & 0.52(7) & 1.58(9) & 1.60(15) & 816.3\,(953) \\
0.705--0.753 & 0.46(6) & 1.55(9) & 1.83(24) & 810.5\,(930) \\
0.753--0.817 & 0.53(5) & 1.69(5) & 1.64(45) & 879.8\,(1017) \\
0.817--0.876 & 0.44(6) & 1.63(9) & 1.11(19) & 862.6\,(958) \\
0.876--1.000 & 0.29(4) & 1.40(7) & 1.03(17) & 1630.5\,(1730) \\\hline
\end{tabular}}
\end{center}
\vspace{-2 mm}
\footnotesize{Note. Numbers in parentheses are 1$\sigma$ uncertainties in the least significant digits.\\
\marka{2--10\,keV flux in units of $10^{-12}$\,\fluxcgs.\\}
}
\end{table}

Given that PL models adequately represent the spectra from individual observatories,
we jointly analyzed the multi-observatory spectra in each phase bin
utilizing common (tied) spectral parameters.
While this approach neglects orbit-to-orbit variability, it provides a clearer picture of
the overall orbital variability of the spectrum.
The results are presented in Table~\ref{ta:ta2} and as orange curves in Figure~\ref{fig:fig2}.
These results broadly agree with the XRT-alone results,
though some discrepancies are evident in $N_{\rm H}$ and $\Gamma$ values, as mentioned above.

\subsection{Spectral Analysis for Two States within the X-ray Peak Interval}
\label{sec:sec2_4}
Our analysis of the light curve revealed a significant change in the flux level,
primarily between two states (see Section~\ref{sec:sec2_2}), in the orbital phase interval of $\phi=0.315\textrm{--}0.365$. To further measure the spectral properties of
these states, we carry out spectral analysis for each of the states.

We constructed separate spectra for the low-flux and high-flux
states within the same orbital phase range ($\phi=0.315\textrm{--}0.365$),
and jointly fit the multi-instrument spectra within each of the states with a PL model.
The low and high states contain 12+2 (XRT+NuSTAR) and 12+1 (XRT+Chandra) spectra, respectively.
For the fit of spectra within each state, we used common $N_{\rm H}$ and $\Gamma$, but separate flux
for each of the spectra.

\begin{figure}
\centering
\includegraphics[width=3.3 in]{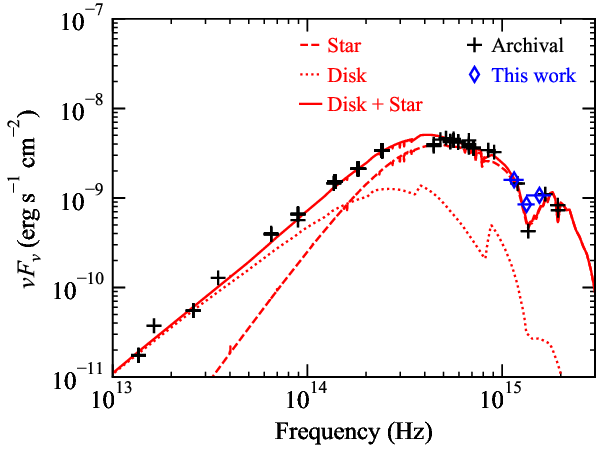} 
\figcaption{
IR-to-Optical SED of J0632. Red lines represent emission components: stellar atmosphere (dashed; see text), disk \citep[dotted;][]{Kim+2022}, and their sum (solid). Our UVOT measurements are shown in blue,
with a stellar atmospheric model including
interstellar extinction overlaid in red (see text). Black points represent
archival measurements \citep[taken from][]{Aragona+2010, Kim+2022}.
\label{fig:fig3}}
\vspace{0mm}
\end{figure}

The derived flux values for the low and high states reveal
variations within each state, characterized by standard deviations of approximately 5--10\%; these variations are statistically insignificant.
The best-fit values for $N_{\rm H}$ and $\Gamma$ for the low- ($N_{\rm H}=(0.73\pm0.06)\times 10^{22}\rm \ cm^{-2}$ and $\Gamma=1.71\pm0.05$)
and high-state ($N_{\rm H}=(0.69\pm0.04)\times 10^{22}\rm \ cm^{-2}$ and $\Gamma=1.67\pm0.03$) spectra
were statistically indistinguishable. The measured 2--10\,keV fluxes ($F_{2-10\rm keV}$) were
$(2.00\pm 0.23)\times 10^{-12}$\,\fluxcgs\ and $(2.92\pm0.22)\times 10^{-12}$\,\fluxcgs\ for the low and high states, respectively.

\section{Analysis of Optical Observations}
\label{sec:opt}
We analyzed optical photometric data from Swift/UVOT and spectroscopic data from the Southern African Large Telescope (SALT).

\subsection{Swift/UVOT Photometric Data}
\label{sec:optphto}
The Swift/UVOT observations were conducted simultaneously with XRT observations (Table~\ref{ta:ta1})
primarily using the UVW1, UVM2, and UVW2 filters. Due to the extreme brightness of J0632's companion exceeding
UVOT's nominal dynamic range, direct photometry from heavily saturated images is unreliable.
We therefore employed the read-out-streak photometry technique developed by \citet{Page+13},
calibrated to bright-limit Vega magnitudes of 8.86, 8.27, and 8.80 for UVW1, UVM2, and UVW2,
respectively. While U-band observations were also obtained, the source brightness exceeded
the capabilities of even this technique.

\begin{figure}
\centering
\hspace{-5mm}
\includegraphics[width=3.36 in]{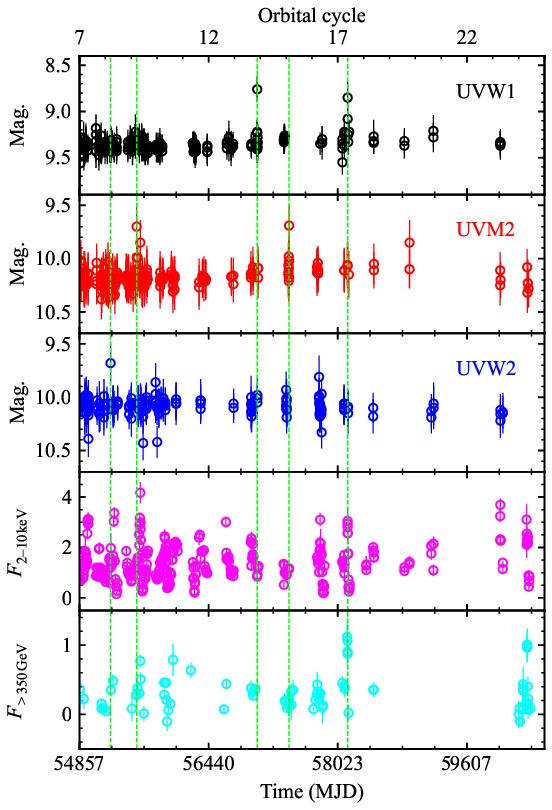}
\figcaption{
Unfolded UVOT light curves in UVW1, UVM2, and UVW2 bands (three panels from top).
X-ray and TeV light curves, in units of $10^{-12}$\,\fluxcgs\ and $10^{-11}$\,\fluxcgs, respectively, (Section~\ref{sec:sec3_1}) are included in the bottom two panels for comparison. Vertical lines denote epochs of UV flares.
\label{fig:fig3x}}
\vspace{0mm}
\end{figure}

Following \citet{Page+13}, we masked bright pixels (e.g., contaminating sources)
and extracted source counts from a 16-pixel-wide column along the read-out streak in each raw
image. Background subtraction utilized 128 columns surrounding the source region.
Measured count rates were corrected for coincidence loss and sensitivity effects,
and converted to Vega magnitudes and flux densities using the calibration in \citet{Page+13}.
The resulting magnitudes (9--11) fall within the validity range of this method. Figures~\ref{fig:fig3} and \ref{fig:fig3x} present the results of our UVOT data analysis.

We compared the measured UV-band flux densities to a stellar atmospheric model.
Measurements deviating by more than 3$\sigma$ from the average (i.e., flares) were excluded.
For the atmospheric emission, we used the TULSTY BSTAR2006 model \citep[][]{Lanz+07}
assuming a surface temperature of $3\times 10^{4}$\,K, a surface gravity of $\mathrm{log}g=4.0$, and
solar metallicity \citep[e.g., see][]{Aragona+2010}.
This stellar atmosphere model was then combined with the disk emission model of \citet{Kim+2022}. The combined model, binned at $50$\AA{} and corrected for interstellar extinction \citep[$A_V=2.4$;][]{Zhu+17}, was normalized to match archival optical data (black points in Figure~\ref{fig:fig3}).
Our UV measurements (blue in Figure~\ref{fig:fig3}) show good agreement with the model,
indicating that the UV emission originates from the companion star.

We then constructed UV light curves. While generally stable,
sudden flux increases by 40\%--70\% are evident in Figure~\ref{fig:fig3x}, likely attributed to
stellar activity such as flares (see above).
These UV flares do not exhibit a significant correlation with X-ray or TeV flares (Figure~\ref{fig:fig3x}), although the optical and TeV flares in cycle 17 appear to occur contemporaneously.
To investigate orbital modulation, folded light curves (excluding the flares) were generated
revealing no significant orbital modulation in the UV band.

\begin{figure}
\centering
\includegraphics[width=\columnwidth]{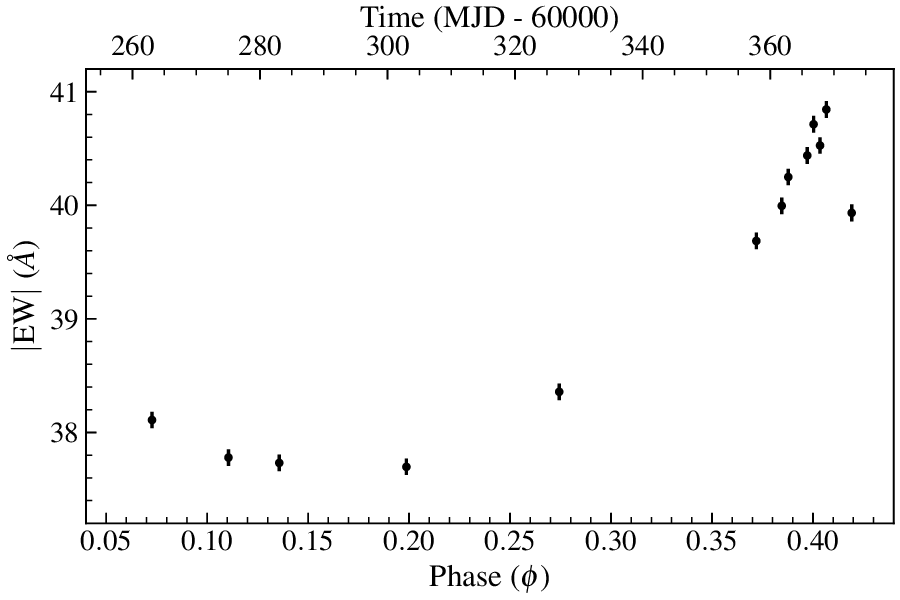}
\figcaption{The H\,$\alpha$ equivalent width measured during orbit 24.\label{fig:salt_halpha}}
\end{figure}

\subsection{Spectroscopic Data from SALT}
\label{sec:optspec}

\begin{figure*}
\centering
\begin{tabular}{cc}
\hspace{-4 mm}
\includegraphics[width=1.9 in]{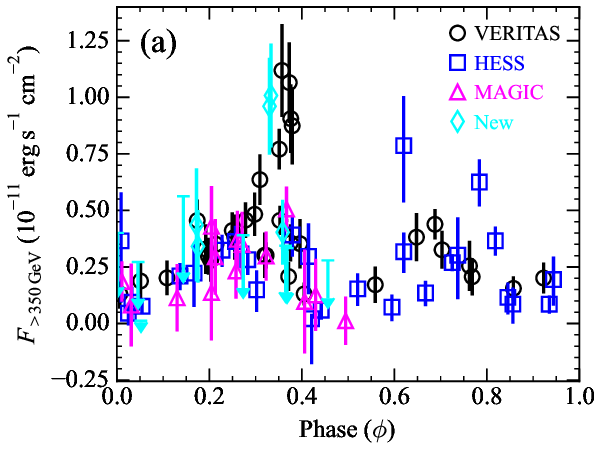} &
\hspace{-4 mm}
\includegraphics[width=5.18 in]{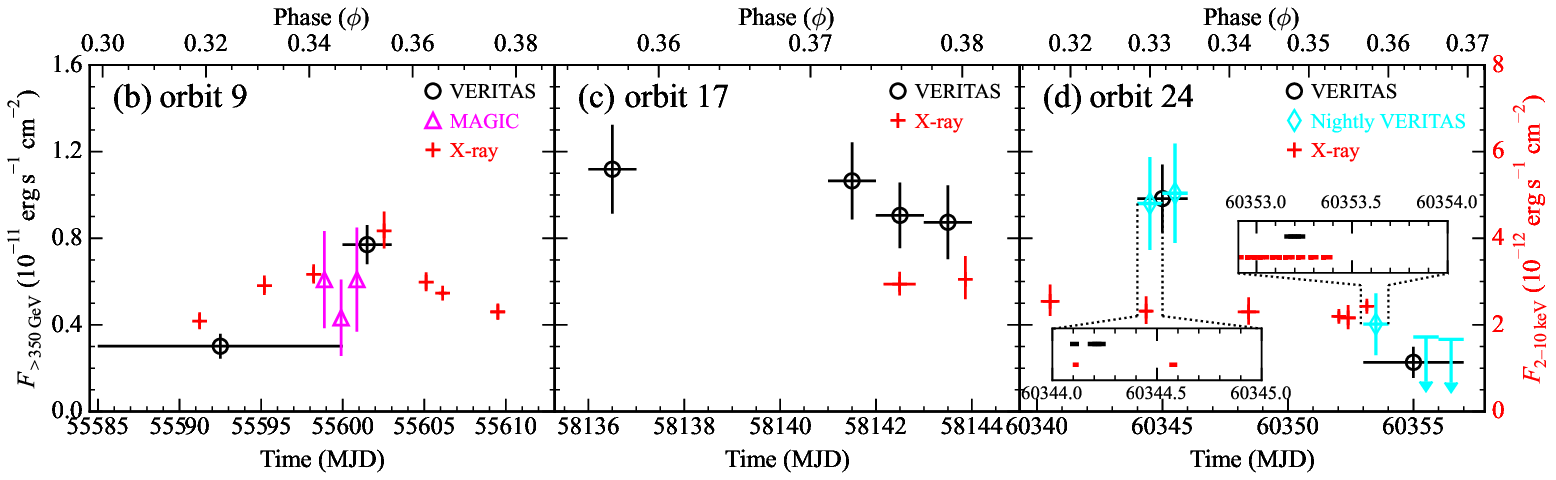} \\
\end{tabular}
\figcaption{(a) Folded light curve of TeV gamma-ray emission ($>350$\,GeV) from J0632 measured by VERITAS, H.E.S.S, and MAGIC \citep[see also Figure~9 of][]{Adams+2021}. New VERITAS measurements from orbital cycle 24 are presented in cyan. (b--d) Unfolded light curves for orbital cycles 9, 17, and 24, respectively. Orbit 9 includes measurements obtained with MAGIC (magenta) and VERITAS (black). In each panel, black circles represent TeV flux measurements (left axis), and red crosses show the 2--10 keV X-ray flux (right axis). In panel (d), cyan diamonds represent daily TeV fluxes, and black circles show the results of an analysis using coarser time bins (see text). TeV fluxes with detection significances below $2\sigma$ are shown as 95\% upper flux limits. Insets provide a visual representation of the X-ray (red) and TeV (black) observation periods to demonstrate overlapping exposures in orbit 24.
\label{fig:fig4}}
\vspace{0mm}
\end{figure*}
Observations were undertaken with SALT \citep{Buckley2006} on 13 nights between 2023 November and 2024 March using the High Resolution Spectrograph \citep[HRS;][]{Bramall2010} in High Resolution (HR) mode, which provides a resolving power of $R\sim65\,000$ over the optical wavelength range 3700–5550\,\AA{} (blue arm) and 5550–8900\,\AA{} (red arm). Each observation consisted of a 3$\times$700\,s exposure.  All CCD reductions, wavelength calibrations, and spectral extractions of the different echelle orders were performed using the SALT HRS pipeline \citep{Kniazev2016}. Individual orders were then merged, barycentric correction was performed, and nightly spectra were averaged together following the standard procedures with {\sc iraf/pyraf}. The EW of the H\,$\alpha$ emission line, which arises from the Be star's circumstellar disk, was measured by performing a summation under the line, with the uncertainty and S/N of the spectrum calculated following \citet{Vollmann2006} and \citet{Stoehr2008}, respectively.

The EWs found before phase $\phi=0.3$ are consistent with recent observations \citep{Matchett+2025}, but a rapid increase is seen after phase $\phi \sim 0.37$, with a decrease around $\phi \approx 0.42$ (Figure~\ref{fig:salt_halpha}).  This would be consistent with increased tidal forces close to periastron. A similar result is observed for the other Be gamma-ray binary system, PSR~B1259$-$63, which shows orbital modulation of the H\,$\alpha$ emission line, where the EW peaks after periastron \citep[see][and references therein]{Chernyakova2025}. These optical results are consistent with the interpretation that the peak in the X-ray and TeV light curve ($\phi \approx 0.35$) occurs around the phase of the disk crossing.

\section{Analysis of VHE Gamma-Ray observations}
\label{sec:sec3}
VERITAS (Very Energetic Radiation Imaging Telescope Array System) is an array of four imaging atmospheric Cherenkov telescopes situated at the Fred Lawrence Whipple Observatory in Arizona, USA \citep{Weekes_2002}. 
Each telescope has a reflector with a diameter of $12\ \mathrm{m}$ and a camera comprising 499 photomultiplier tube pixels, providing a field of view of $3.5^\circ$. 
VERITAS is sensitive to photons with energies ranging from $85\ \mathrm{GeV}$ to $100\ \mathrm{TeV}$, achieving an angular resolution of better than $0.08^\circ$ (68\% containment radius) above $1 \ \mathrm{TeV}$. 
The observations used in this paper span from 2023 October to 2024 March, corresponding to orbit 24. Following data quality selection, a total of $17\,\mathrm{h}$ effective exposure is available for point source analysis at the position of the X-ray source XMMU J063259.3+054801.

\subsection{TeV Light Curve}
\label{sec:sec3_1}

We fit the daily spectra with a PL model, holding $\Gamma$ fixed at $2.6$ \citep{Adams+2021},
to derive the source flux (cyan in Figure~\ref{fig:fig4}a).
Five daily observations taken between 2023 October and 2024 March were conducted around the X-ray peak phase interval of $\approx 0.33$.
The source was bright and well detected during the first two observation nights, see Figure~\ref{fig:fig4}d.
The fit-inferred flux values above 350\,GeV ($F_{\rm >350\ GeV}$) for these two spectra were consistent.
The other three observations were made over a 4-day period, approximately 8\,days after the first two.
The source was fainter during this period, and a significant detection was achieved only
in the first of these three observations. It seems that the source flux decreased
in the next two days, and the source was not significantly detected.
Outside the X-ray peak phase interval, the source was not significantly detected (significance $<2\sigma$) after MJD 60355.

To improve the flux measurements (and source detection),
we used coarse time bins. We combined spectra collected within 2--4 days,
resulting in 8 combined spectra. While the overall results were not significantly affected
(compared to the analysis of the daily data),
the uncertainties in the flux measurements were slightly reduced.

\begin{table}[t]
\vspace{-0.0in}
\begin{center}
\caption{Gamma-ray flux measurements by VERITAS around the X-ray peak of orbit 24}
\label{ta:ta3}
\vspace{-0.05in}
\scriptsize{
\begin{tabular}{cccc} \hline
MJD Range &  Exposure & Flux\marka & Significance \\
 &  (h) & ($10^{-12}\rm \ s^{-1}\ cm^{-2}$) &  \\ \hline
\multicolumn{4}{c}{Nightly averages} \\
60344.09--60344.24 & 2.0 &  $6.0\pm1.3$ &  6.0 \\
60345.17--60345.24 & 1.5 & $6.3\pm1.4$ & 6.2 \\
60353.15--60353.24 & 2.0 & $2.5\pm0.9$ & 3.5 \\
60355.14--60355.23 & 2.0 & $<2.14$\markb & 1.9 \\
60356.14--60357.19 & 1.0 & $<2.07$\markb & 0.5 \\ \hline
\multicolumn{4}{c}{Averages over several nights} \\
60344.09--60345.24 & 3.6 & $6.1 \pm 1.0$ & 8.6 \\
60353.15--60357.19 & 5.0 & $1.4 \pm 0.6$ & 3.8 \\ \hline
\end{tabular}}
\end{center}
\vspace{-2 mm}
\footnotesize{
\marka{Fluxes are determined for energies $>350$ GeV assuming a spectral index of $2.6$.}\\
\markb{Upper flux limits at 95\% confidence level are given for measurements with detection significance below $2\sigma$.}\\}
\end{table}

We constructed a $>350$\,GeV light curve by combining our analysis with previous
measurements reported by \citet{Adams+2021}.
The light curve is displayed in Figure~\ref{fig:fig3x} and Figure~\ref{fig:fig4}a.
Prominent TeV flares (e.g., $F_{\rm >350\ GeV} > 6\times 10^{-12}$\,\fluxcgs; see also Section~\ref{sec:sec3_2}) were observed around the X-ray peak phase interval in orbital cycles 9, 17, and 24 (Figure~\ref{fig:fig4}b--d). To measure the timescale of TeV flux variability within this phase interval, we inspected the unfolded light curve; substantial flux changes ($2\sigma$) were observed on timescales of $\lapp5\textrm{--}8$\,days. 

Simultaneous X-ray observations were obtained during the TeV-flaring epochs corresponding to orbital cycles 9, 17, and 24.  The X-ray and TeV light curves for these cycles are shown in Figure~\ref{fig:fig4} (b)--(d). TeV fluxes are from \citet{Adams+2021}; new measurements for orbit 24 are listed in Table \ref{ta:ta3}. A simultaneous increase in the X-ray and TeV fluxes is evident during cycle 9 (Figure~\ref{fig:fig4}b). In contrast, they appeared to vary independently during cycle 24 (Figure~\ref{fig:fig4}d). UV flares, as described in Section~\ref{sec:optphto}, were observed coincident with the TeV flares in cycle 17, but no UV flares were detected during cycles 9 and 24. While these UV/X-ray and TeV observations are contemporaneous on a daily timescale, they do not perfectly overlap, as shown in the insets of Figure~\ref{fig:fig4}d. Thus, the observed variabilities could be influenced by very short-timescale fluctuations, on the order of hours.

\subsection{Correlation between the X-ray and TeV Flux Variabilities}
\label{sec:sec3_2}

To investigate the correlation between the X-ray and TeV fluxes reported by \citet{Adams+2021},
we performed spectral analysis on individual Swift/XRT spectra.
Due to limited counts and parameter covariance in spectral fits of the XRT data, we fixed $N_{\rm H}$ and $\Gamma$ at the values determined from the joint analysis (Table~\ref{ta:ta2}) of data from the corresponding phase interval when deriving the X-ray flux.
We then created a plot of TeV flux versus X-ray flux, shown in Figure~\ref{fig:fig5}  (including both new and archival data).
For this, we utilized contemporaneous X-ray and TeV datasets where the X-ray observation(s)
were conducted within the time range \citep[e.g., Table~\ref{ta:ta3} and][]{Adams+2021} of the corresponding TeV observation. This temporal constraint was imposed due to the known variability of the J0632 X-ray flux on a $\sim1$-day timescale (Section~\ref{sec:sec2_2}).

We conducted a correlation analysis between X-ray and TeV flux measurements using Pearson's correlation coefficient ($r_{\rm P}$).
The calculated $r_{\rm P}$ is 0.744. Initially, we estimated the probability of observing such a correlation from uncorrelated variables to be $p\approx 6\times 10^{-8}$ (corresponding to $5.3\sigma$) using the method described in \citet{Bevington2003}.
However, recognizing that this method assumes Gaussian distributions for the measured flux values and does not account for measurement uncertainties, we performed simulations to address these limitations. For each observed flux, we generated simulated data by varying the measured values according to their uncertainties. We then constructed distributions for both the simulated X-ray and TeV data. From these distributions, we generated 10,000 pairs of uncorrelated datasets and computed $r_{\rm P}$ for each pair. 
The resulting distribution of $r_{\rm P}$ values has a mean of 0 and a standard deviation of 0.164. Comparing our observed correlation coefficient of 0.744 to this simulated distribution, we find its significance to be $4.6\sigma$. This confirms the correlation previously reported by \citet{Adams+2021}.

We further quantified the observed correlation trend by fitting it with a linear function, $F_{>\rm 350 GeV}=a F_X + b$. For this fitting procedure, we employed the $\bar \eta^2$ method \citep{B1259HESS2024}, which accounts for uncertainties in both $F_X$ and $F_{>\rm 350 GeV}$. The model effectively describes the data, yielding a low $\bar \eta^2=7.3$. The probability of achieving an $\bar \eta^2$ value lower than the observed was estimated to be 0.05 through simulations. The best-fit parameters were found to be $a=2.66\pm0.23$ and $b=(-1.16\pm0.37)\times 10^{-12}$\,\fluxcgs\ (Figure~\ref{fig:fig5}).

To examine the TeV flare phase interval ($\phi=0.3$--$0.4$; Figure~\ref{fig:fig4}),
we highlighted the data points within this phase interval in blue in Figure~\ref{fig:fig5}.
These data points appear to show high and low TeV states
around a dividing TeV flux of $6\times 10^{-12}$\,\fluxcgs.
Low-flux points within the TeV flare interval follow a trend similar to that seen in other phase intervals.
Although high-flux TeV points also generally follow this trend, they exhibit significant scatter.
Moreover, these high-flux points alone do not demonstrate a clear positive or negative correlation with X-ray variability;
the TeV flux does not vary much despite large X-ray fluctuations.

\begin{figure}
\centering
\includegraphics[width=3.1 in]{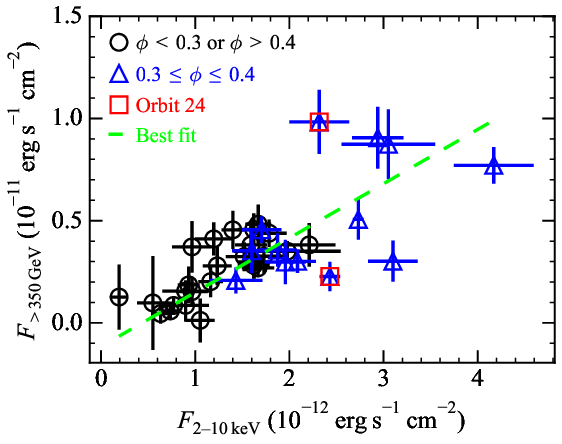}
\figcaption{TeV flux ($>350$\,GeV) vs. 2--10\,keV X-ray flux for contemporaneous observations of J0632. The green dashed line represents the best-fit linear function derived from $\bar \eta^2$ minimization (see text for more detail). \label{fig:fig5}}
\end{figure}

\section{Discussion}
\label{sec:sec4}
We analyzed X-ray, optical, and TeV observations of the TGB J0632, incorporating data from 24 orbital cycles.
By aligning the phases of new Swift/XRT data with archival ones, we measured the orbital period to be
$316.65\pm 0.31$\,days (Section~\ref{sec:sec2_2}), consistent with previous comprehensive X-ray and gamma-ray studies \citep[][]{Adams+2021}.

The increased data volume allowed for a more detailed characterization
of the previously reported orbital variability in the source's X-ray spectrum.
While our measurements of the orbital variations in flux, $N_{\rm H}$, and $\Gamma$ are largely
consistent with those reported by \citet{Malyshev+2019},
the new analysis provided a more detailed spectral characterization, including a determination of the phase width of the flux and $N_{\rm H}$ increase as $\Delta \phi\approx0.1$.

We found that the source exhibits two X-ray flux states within the phase interval $\phi=0.315$--$0.365$ and that transitions between the states occur on both long (orbital period) and short timescales ($\lapp 3$\,days).
Our multi-wavelength campaign revealed that
the source's X-ray and TeV variability in orbital cycle 24 appear independent,
contrary to the previously-observed concurrent flux increase (e.g., orbit 9) in these two bands.

Our analysis of previously unreported Swift/UVOT data reveals that the UV emission from the J0632 companion star is generally stable, though short-term flares are observed. Notably, these UV flares are observed across a wide range of orbital phases, in contrast to the X-ray and TeV flares, which are primarily associated with the interaction phase.

Below, we discuss these observations in the context of IBS emission and pulsar-disk interaction.

\subsection{Orbital Variability of the X-ray Spectrum}
\label{sec:sec4_1}
As mentioned previously, some of the observed orbital variability in the X-ray spectrum
might be attributed to parameter covariance and orbit-to-orbit variability in the source's emission,
particularly given the multi-orbit nature of the Swift observations.
However, certain features, such as the increased $N_{\rm H}$ and flux at $\phi\approx 0.3\textrm{--}0.4$, are highly significant and
are confirmed by single-epoch data from other X-ray observatories \citep[e.g.,][]{Malyshev+2019,Kargaltsev+2022}.
The observed increases in the X-ray flux and $N_{\rm H}$ in the phase interval have been attributed to the presence
of a circumstellar disk. These variabilities can potentially be used to constrain
the geometry of the binary system and the location of the Be disk.

Previous studies have proposed different disk configurations based on the $N_{\rm H}$ distribution.
\citet{Malyshev+2019} suggested a scenario where the Be disk crosses the orbit at phases
$\phi\approx0.35$ and 0.7, while \citet{Tokayer+2021} proposed disk crossings at $\phi\approx 0.13$ and 0.35,
based on increased X-ray flux and $N_{\rm H}$ around $\phi\approx0.15\textrm{--}0.2$.
While both models readily explain the prominent rise in $N_{\rm H}$ around $\phi=0.35$ as due to high
inclination of the disk, the other phase interval suggested for the disk crossing
could not be confirmed with the existing data, since the increase in $N_{\rm H}$ at $\phi\approx 0.7$
or $0.15$ may only represent an orbit-to-orbit spectral variability and not an intrinsically
higher $N_{\rm H}$. Recently, \citet{Matchett+2025}, based on an analysis of archival and new H\,$\alpha$ data,
suggested a disk configuration similar to that of \citet{Malyshev+2019}, lending
credence to a disk crossing at $\phi\approx0.7$.

Nonetheless, the increase in both the X-ray flux and $N_{\rm H}$ around $\phi=0.35$ is
highly significant. If this increase is indeed caused by disk material,
the phase width of $\approx 0.1$ (Figure~\ref{fig:fig2}a and b) may correspond to the disk opening angle.
While the system's orbital parameters remain undetermined, assuming a circular orbit,
the measured phase width suggests an opening angle of $\approx 36^\circ$; the exact value
depends also on the angle between the orbital plane and the disk.
Notably, this inferred opening angle is comparable to the $37^\circ$ estimated for PSR~B1259-63 based on X-ray and TeV observations \citep[][]{Chernyakova2006}.

On the other hand, these $N_{\rm H}$-driven orbital solutions \citep[][]{Malyshev+2019,Tokayer+2021} are incompatible
with ones proposed by \citet{Casares+2012} and \citet{Moritani+2018} \citep[see also ][]{Matchett+2025}
based on radial velocity measurements
(we should note that these radial-velocity-inferred orbits are incompatible with each other).
Conversely, the observed variabilities in the X-ray flux and $N_{\rm H}$ pose a challenge to
these orbital configurations,
as the Be disk in these models crosses the orbit at different phases from those having increased $N_{\rm H}$.
This necessitates exploring alternative explanations for the flux and $N_{\rm H}$
increases at $\phi\approx 0.35$ (and potentially at $\phi\approx0.15$ or $\phi\approx0.75$).
One potential explanation, proposed by \citet{Moritani+2018}, is that the source undergoes an outburst
when it approaches the misaligned disk, which might account for the simultaneous increase in flux and $N_{\rm H}$.
However, the authors noted that this interpretation struggles to explain the observed difference
in the amplitudes of the flux enhancements, with the flux at $\phi\approx 0.35$ being significantly
larger than the one at $\phi\approx 0.7$.

\subsection{X-ray Flux Variability within the Peak Phase Interval}
\label{sec:sec4_2}
During the X-ray flare phases ($\phi=0.315\textrm{--}0.365$), the source exhibited apparent low- and high-flux states,
characterized by Swift/XRT count rates $<0.055$\,cps and $\gapp 0.06$\,cps, respectively (Section~\ref{sec:sec2_2}).
Given the recurrent nature of this high-flux state (X-ray flare) within the specific orbital phase interval,
it is likely attributed to a stationary structure in the system, such as the companion's disk.
This suggests that pulsar-disk interactions may occur during this phase interval.
In this case, a stronger momentum flux from the decretion disk may be responsible for the
increased X-ray flux as the disk may compress the IBS towards
the pulsar, leading to an increased $B$ within the emission region \citep[][]{Chen2019,Kim+2022}.

In this IBS scenario, our observations of the long-term change in the X-ray flux
at the peak phase interval (Figure~\ref{fig:fig1}c)
suggest that the IBS was pushed closer to the pulsar
by the stronger momentum flux of the disk prior to MJD~55800,
which subsequently decreased substantially over a timescale of the orbital period.
This is consistent with observations of H\,$\alpha$ line emission, which
showed a substantial decrease in the EW between MJDs 55700 \citep[$\gapp 50$\AA{};][]{Casares+2012} and 56600 \citep[$\approx 30$\AA{};][]{Moritani+2015},
indicating that the disk was larger at the earlier epoch \citep[][]{Grundstrom+2006}. Weaker X-ray emission (Figure~\ref{fig:fig1}c) and smaller H\,$\alpha$ EW \citep[$\sim 30$\AA{};][]{Adams+2021} in cycle 16 ($\sim$MJD~57800) are also consistent with this picture. In orbit 24, H\,$\alpha$ EWs were measured to be $38$--$41$\AA{} (Section~\ref{sec:optspec}), and J0632 was in the low-X-ray-flux state during the interaction phase.
The mechanism by which this scenario (disk compression) leads to the apparent ``two'' X-ray flux states remains unclear.  It is possible that the source exhibits a continuous range of X-ray flux states during the interaction phase, and our classification is a simplification of this continuum.

It is important to note that \citet{Tokayer+2021} reported larger H\,$\alpha$ EW values
between MJDs 58700 and 58900 ($\sim50$\AA{}), while \citet{Matchett+2025}
measured smaller values between MJDs 59200 and 60000 ($36$--$40$\AA{}).
These measurements suggest the presence of a larger disk at the earlier period
and a smaller disk at the later period, respectively.
The companion's disk appears to undergo relatively frequent changes.
Further simultaneous optical and X-ray observations can clarify the
relationship between H\,$\alpha$ EW and the X-ray flux state during the interaction phase.

Short-term X-ray variability is a common feature of TGBs \citep[e.g.,][]{Bosch+05,Smith+09}
and has been attributed to clumpy structures within the stellar wind. \citet{Kefala+23} investigated
the impact of clumpy winds on the deformation of IBSs.
Such deformations can alter the location and size of the emission volume,
resulting in changes in the X-ray emission. It was demonstrated that X-ray variability
on diverse timescales can arise, depending on the size of the clump.

In this context, the observed short-term variability in J0632's X-ray emission
during the interaction phase may imply a clumpy structure in the disk.
Additionally, the observed 1-day \citep[or shorter;][]{Kargaltsev+2022} variability at various orbital phases (outside the interaction phase)
can be attributed to the clumpy wind of the companion.
These variability timescales (Section~\ref{sec:sec2_2})
can provide insights into the size of these clumps,
provided that the orbital speed of the pulsar (i.e., system orbit) is known.
We caution that the measured variability timescales are limited by the
cadence of the Swift observations; thus, the values should be regarded as upper limits.

\subsection{Variability in the X-ray and TeV Fluxes}
\label{sec:sec4_3}
Using the new and archival X-ray and TeV data, we confirmed the correlation reported by \citet{Adams+2021}
between the X-ray and TeV flux variabilities across the entire orbital phase interval (Figure~\ref{fig:fig5}).
A similar correlation has also been reported for PSR~B1259$-$63 \citep[][]{B1259HESS2024}.

However, our contemporaneous X-ray and TeV observations of J0632 within the interaction phase ($\phi\approx0.3$--$0.4$)
in orbital cycle 24 revealed independent variations in TeV and X-ray fluxes (Section~\ref{sec:sec3_1}).
Specifically, within $\lesssim 8$\,days during the flare phase, the TeV flux decreased while the
X-ray flux remained constant (Figure~\ref{fig:fig4}d).
It is difficult to identify a similar trend within the disk-crossing phase intervals
of the archetypal TGB PSR~B1259$-$63 in the results of \citet{B1259HESS2024} because the phase width
for the disk crossing is not well specified, and contemporaneous X-ray and TeV measurements
are lacking around the interaction phase for this source.
Therefore, J0632 may offer a unique opportunity to probe the radiation mechanisms
during the interaction phase.

IBS models of TGBs \citep[e.g.,][]{d06,ar17} attribute their X-ray emission to synchrotron radiation
from electrons within the shock and the TeV emission to IC scattering of these electrons
off seed photons from the companion and its disk.
Since the same electrons are responsible for both X-ray and TeV photons,
the ratio of X-ray and TeV flux is determined by $u_B/(u_* + u_d)$, where
$u_B$ is the magnetic energy density within the shock, and 
$u_*$ and $u_d$ are the energy densities of the seed photon field supplied by
the companion and its disk, respectively. 
The UV flares observed in orbit 17 suggest an increase in $u_*$, which could contribute to the observed increase in TeV flux during that cycle.

The absence of UV flares in orbits 9 and 24 suggests that the observed TeV enhancements in these orbits are due to increased disk emission.
An increase in $u_d$ at the disk-crossing phase is expected in the
model proposed by \citet{vanSoelen+2012} and \citet{Chen2019},
where shock heating during the pulsar-disk interaction phase can increase $u_d$.
Enhanced disk emission then provides seed photons for IC scattering by the IBS electrons,
boosting TeV emission. This model predicts increases in both $u_B$ and $u_d$
at the interaction phase \citep[see][]{Chen2019}, consistent with the observations
in orbit 9 (Figure~\ref{fig:fig4}b).

A key challenge to the model is the presence of TeV-only flares in orbit 24, characterized by high TeV fluxes and low X-ray fluxes (Figure~\ref{fig:fig4}d).
The rapid TeV variability observed during this cycle, with significant flux changes on a timescale of $\lesssim 8$ days despite relatively stable X-ray emission, further complicates the picture.  The disk heating scenario proposed by \citet{Chen2019} is unlikely to be responsible for these events, 
as it predicts concurrent increases in both $u_B$ and $u_d$ (see above), which should be accompanied by corresponding increases in X-ray flux.  Given the absence of a UV flare (indicating no increase in $u_*$), it is probable that an increase in $u_d$ is responsible for the TeV-only flares.  However, the physical mechanism capable of increasing $u_d$ without also increasing $u_B$ remains an open question.

As we mention in Section~\ref{sec:sec4_2},
the source might have undergone a long-term change in its disk
between orbits 9 and 10 (Figure~\ref{fig:fig1}c; and between other orbital cycles as noted in Section~\ref{sec:sec4_2}).
Prior to orbit 9, the companion's disk was extended, perhaps reaching the pulsar orbit,
as evidenced by H\,$\alpha$ measurements.
This could explain the effectiveness of the disk heating scenario of \citet{Chen2019} in orbit 9,
contrasting with its inapplicability in orbit 24 (i.e., smaller disk).

However, the observed independent variability in the X-ray and TeV fluxes could be attributed to very short-timescale ($\sim \textrm{hours}$) variabilities not fully captured by the existing data, as noted previously (Section~\ref{sec:sec3_1}). Therefore, a definitive conclusion cannot be drawn based solely
on these observations.
Additional simultaneous optical, X-ray, and TeV observations could illuminate the connection
between disk size and high-energy emission, offering deeper insights into the underlying
microphysics of the interaction and shock heating. Developing theoretical models that incorporate the microphysical details of this interaction is essential
for a comprehensive interpretation of these observational data.

\section{Summary}
\label{sec:sec5}
Our analysis of multi-wavelength observations targeting HESS~J0632+057 has revealed complex phenomena.
Below is a summary of our work.
\begin{itemize}
\item The observed increases in flux and $N_{\rm H}$ around $\phi=0.35$ provide constraints on
the configuration of the companion's disk, suggesting the pulsar's passage through the disk during this interval. This interpretation is further supported by the indication of variability of the H\,$\alpha$ EW around these orbital phases.
\item The measured phase widths of $\Delta \phi\approx 0.1$ for the flux and $N_{\rm H}$ enhancements imply a
disk opening angle of $\approx 36^\circ$ for an assumed circular orbit.

\item The X-ray flux during the interaction phase ($\phi=0.315\textrm{--}0.365$; Section~\ref{sec:sec2_2}) exhibited both short- and long-term variability.
Long-term variability may indicate changes in the disk size,
while short-term variability may be caused by clumpiness of the disk material.

\item The TeV flares observed from J0632 within the interaction phase appear to be
driven by multiple mechanisms.
Discerning between interaction-induced flares and those arising from other effects is
crucial for understanding pulsar-disk interactions. Simultaneous optical, X-ray and TeV
observations are essential to achieve this goal.
\end{itemize}

\begin{acknowledgments}
This work used data from the NuSTAR mission, a project led by the California Institute of Technology,
managed by the Jet Propulsion Laboratory, and funded by NASA. We made use of the NuSTAR Data
Analysis Software (NuSTARDAS) jointly developed by the ASI Science Data Center (ASDC, Italy)
and the California Institute of Technology (USA).

This paper employs a list of Chandra datasets, obtained by the Chandra X-ray Observatory, contained in Chandra Data Collection\dataset[DOI: 10.25574/cdc.345]{https://doi.org/10.25574/cdc.345}.

This research is supported by grants from the US Department of Energy Office of Science, the US National Science Foundation, and the Smithsonian Institution, by NSERC in Canada, and by the Helmholtz Association in Germany. 
This research used resources provided by the Open Science Grid, which is supported by the National Science Foundation and the U.S. Department of Energy's Office of Science, and resources of the National Energy Research Scientific Computing Center (NERSC), a U.S. Department of Energy Office of Science User Facility operated under Contract No. DE-AC02-05CH11231. 
The authors acknowledge the excellent work of the technical support staff at the Fred Lawrence Whipple Observatory and at the collaborating institutions in the construction and operation of the instrument.

Some of the observations reported in this paper were obtained with the Southern African Large Telescope (SALT) under program 2020-2-MLT-007 and 2023-2-DDT-004 (PI: B.\ van Soelen).
JP acknowledges support from the Basic Science Research Program through the National
Research Foundation of Korea (NRF) funded by the Ministry of Education (RS-2023-00274559).
This work was supported by the National Research Foundation of Korea (NRF)
grant funded by the Korean Government (MSIT) (RS-2023-NR076359). This work was conducted during the research year of Chungbuk National University in 2025. Support for this work and KM was provided by NASA through NNH22ZDA001N-NICER. BvS acknowledges support from the National Research Foundation of South Africa (grant number 119430).
We thank the referee for comments that helped improve the clarity of the paper.
\end{acknowledgments}

\facilities{SALT, Swift, Suzaku, XMM, CXO, NuSTAR, VERITAS}
\software{
XSPEC \citep[v12.14.0;][]{a96},
HEAsoft \citep[v6.33;][]{heasarc2014},
CIAO \citep[v4.16;][]{CIAO2013},
XMM-SAS \citep[20230412\_1735;][]{xmmsas14},
Eventdisplay \citep[v490.7;][]{maier_2024_14283930}
}

\bibliography{J0632}{}
\bibliographystyle{aasjournalv7}
\end{document}